# Quadrupolar ordering and exotic magnetocaloric effect in $R$B$_4$ ($R$ = Dy, Ho)

M.S. Song[1], K. K. Cho[1], B.Y. Kang[1], S. B. Lee[2], B. K. Cho[1*]

[1]*School of Materials Science and Engineering, Gwangju Institute of Science and Technology (GIST), Gwangju 61005, Korea*

[2]*Department of Physics, Korea Advanced Institute of Science and Technology (KAIST), Daejeon, 34141, Korea*

## ABSTRACT

The interplay of charge, spin, orbital and lattice degrees of freedom has recently received great interest due to its potential to improve the magnetocaloric effect (MCE) for the purpose of magnetic cooling applications. Here we propose a new mechanism for a giant inverse MCE in rare-earth tetraborides, especially for Ho$_{1-x}$Dy$_x$B$_4$ ($x$ = 0.0, 0.5, and 1.0). For $x$ = 0.0, 0.5, and 1.0, the maximum entropy changes of the giant inverse MCE are found to be 22.7 J/kg·K, 19.6 J/kg·K, and 19.0 J/kg·K with critical fields of ≈ 25 kOe, 40 kOe, and 50 kOe, respectively. It is remarkable that such a giant MCE is realized, even when applying a low magnetic field, which enables a field-tuned entropy change and brings about a significant advantage for several applications. For all compounds, we have systematically studied how the entropy changes as a function of the field and temperature and investigated their correlation with consecutive double transitions, i.e., the magnetic dipolar order at $T = T_N$ and the quadrupolar order at $T = T_Q$ ($T_Q < T_N$). We found that the maximum entropy change occurs at $T = T_Q$ and the critical field associated with the meta-magnetic transition, which is in good agreement with the experimental data. Thus, we elucidate that this unique behaviour is attributed to the strong coupling between magnetic dipoles and quadrupoles in the presence of strong spin-orbit coupling and geometric frustration. Our work offers new insights into both the academic interest of multipolar degrees of freedom in magnetic materials and the discovery of giant MCE with various applications for magnetic cooling systems.

I. INTRODUCTION

The magnetocaloric effect (MCE) is a thermodynamic property, in which heating or cooling occurs in magnetic materials when applying a magnetic field. For the conventional MCE, the cooling mechanism is based on the adiabatic demagnetization process. In contrast, the *inverse* situation can also occur, where the system is cooled via adiabatic magnetization. This is often termed as the *inverse* MCE. Refrigeration based on the conventional or *inverse* MCE is a solid-state cooling application, which is energy efficient, noise-free, and environmentally friendly. Thus, a large MCE is attractive as an alternative to conventional vapor refrigeration [1]. In particular, a large MCE in a low-temperature region is being actively studied for the purpose of gas liquefaction (hydrogen and helium), space technology, and diverse scientific research technologies. Although various methods have been devised for the development of novel solid-state cooling, the design and discovery of new materials that exhibit a large MCE are still important.

Intuitively, a large MCE is expected in materials with a first order magnetic phase transition accompanied with a spontaneous magnetization jump. However, this gives rise to heat loss during the refrigeration cycle due to the hysteresis, irreversibility and the narrow working temperature range. As another promising candidate, a system with geometrical frustration may contain an enormous ground state degeneracy due to competing spin exchange interactions, so a large magnetic entropy change is expected when a magnetic field is applied [2,3]. In addition, more exotic scenarios with multipolar degrees of freedom and their influence on the MCE have been proposed [4-6]. Because such multipolar degrees of freedom are expected to give rise to an anomalous MCE, it would be of great interest to find such systems, thereby allowing us to unveil the nature of the anomalous MCE.

Multipolar degrees of freedom and their importance have garnered significant attention in many correlated electronic systems, such as heavy fermions, frustrated magnets, multiferroics and superconductivity [7-12]. For alloys with heavy 4*d*, 5*d* transition metal ions or rare-earth ions, the spin and orbital degrees of freedoms are strongly entangled, and the system may contain a significant correlation between the spin-orbit coupled multipoles, resulting in their spontaneous ordering. When such multipolar degrees of freedom meet geometrical frustration, their interplay gives rise to multiple magnetic phase transitions, which significantly induces a large MCE, even in the absence of a first order magnetic phase transition. Thus, exploring such multipolar degrees of freedom in a strongly spin-orbit coupled system is definitely required when investigating the potential candidate materials with a large MCE. Not only regarding technological applications, this also deepens our understanding of complicated spin systems in the presence of various competing exchange interactions and their exotic multipolar order, which is, in principle, very challenging to detect. Thus, it is often termed as the 'hidden order' [13-16].

Motivated by the above, we propose a new candidate of material rare-earth tetraborides for a large *inverse* MCE and elucidate their theoretical origin, pursuing possible controllability. A rare-earth tetraboride with a chemical formula $R$B$_4$ ($R$ = rare-earth elements) is a system in which strong spin-orbit coupling and geometrical frustration coexist. In the presence of spin-orbit coupling and a crystalline electric field, the spin states in rare earth ions with valence 3+ split into several doublets or singlets in terms of the total angular momentum J basis. It is known that, depending on rare-earth ions, four low lying energy states are quite well separated from the other excited states, forming pseudo-quartets [17,18]. Furthermore, the lattice structure formed by $R$ ions in the *c* plane is the Shastry-Sutherland lattice (SSL), which is a geometrically frustrated system, exhibiting double magnetic transitions, magnetic dipole ordering at $T = T_{N1}$ and quadrupolar ordering at $T = T_{N2}$ [19]. Since the 1970s, the physical properties and magnetic structure of $R$B$_4$ compounds have been studied [20-24] and, recently, the detailed ground state of $R$B$_4$ ($R$ = Dy, Ho) was re-investigated by resonant X-ray scattering and X-ray and neutron diffraction [17, 25-28]. The two successive magnetic transitions of these compounds are related to the collinear antiferromagnetic transition at $T = T_{N1}$ along the *c*-axis, the quadrupolar ordering at $T = T_{N2}$ and the strong quadrupolar fluctuation between them. In addition, a structural transition from tetragonal to monoclinic is also observed at the quadrupolar ordering temperature, which indicates strong quadrupole-lattice coupling.

The MCEs of high-purity single crystals of $R$B$_4$ ($R$ = Dy and Ho) are investigated by examining the temperature- and field-dependence of magnetization with an applied field along the *c*-axis and the *ab* plane. It is quite remarkable that these materials exhibit a giant *inverse* MCE, i.e., a maximum magnetic entropy change near the quadrupolar ordering of +19.6 J/kg·K, +19.0 J/kg·K, and +22.7 J/kg·K at the critical fields of 50 kOe, 40 kOe, and 25 kOe for DyB$_4$, Dy$_{0.5}$Ho$_{0.5}$B$_4$, and HoB$_4$, respectively. The exotic *inverse* MCE increases with an increasing magnetic field below the critical field but decreases above the critical field. The critical field is the regime where the non-collinear magnetic ground state breaks down, thus indicating that the entropy change is clearly correlated with the quadrupolar ordering.

II. EXPERIMENTAL DETAILS

The single crystals of $R$B$_4$ ($R$ = Dy and Ho) are prepared by a high-temperature metal flux method using the Al flux [29,30]. A stoichiometric mixture of rare earth metals (≥ 99.9%, China Rare Metal Material Co., LTD.) and boron pieces (99.9%, RND Korea) are placed in an alumina crucible (99.8%, Samhwa Ceramic Company) together with the Al (99.999%, RND Korea) flux at a mass ratio of $R$B$_4$: Al = 1: 50. The mixture is placed in a heated tube furnace with an MoSi$_2$ heating element. The furnace is heated at a rate of 300 °C per hour to 1650 °C under a high-purity argon atmosphere after dehydration and cooled slowly at a rate of 4.8 °C per hour to 650 °C. The single crystals are separated from the flux by dissolving the excess Al in NaOH.

The crystal structures are characterized using *X*-ray diffraction measurements (XRD; Rigaku D/MAX-2500 with a Cu target) at room temperature. The XRD data are collected on pulverized single crystals of DyB$_4$ and HoB$_4$. The XRD patterns show a single phase of DyB$_4$ and HoB$_4$ without any observable impurity peaks. The Bragg peak positions are in good agreement with the tetragonal symmetry of the ThB$_4$-type structure and space group *P4/ mbm* (#127) [29,30]. The lattice parameters are determined from LeBail refinements using FULLPROF software. The refined lattice parameters are $a$ = 7.099(5) Å and $c$ = 4.015(9) Å for DyB$_4$ and $a$ = 7.083(8) Å and $c$ = 4.005(1) Å for HoB$_4$. The temperature- and field-dependent magnetizations are measured using a superconducting quantum interference device magnetometer (SQUID; Quantum Design MPMS XL).

III. RESULTS AND DISCUSSION

Figure 1 shows the temperature-dependence of the magnetization divided by the applied magnetic field, parallel and perpendicular to the $c$-axis, namely, $M(T)/H$ with $H$ = 10 kOe, for a single crystal of HoB$_4$. There are two successive magnetic transitions at $T_{N2}$ = 5.7 K and $T_{N1}$ = 7 K for both applied magnetic fields parallel and perpendicular to the $c$-axis. Thus, this can be split into three distinct phases, I ($T > T_{N1}$), II ($T_{N1} > T > T_{N2}$), and III ($T_{N2} > T$) in the low-field region with a decreasing temperature. In phase I, the paramagnetic phase follows the Curies-Weiss law, $M(T)/H = C/(T-\theta)$, where $C = N_0\mu_{eff}^2/3k_B$, $N_0$ is Avogadro's number, $k_B$ is the Boltzmann constant, and the effective magnetic moment, $\mu_{eff}$, is determined to be 10.4$\mu_B$, where $\mu_B$ is the Bohr magneton, and the Weiss temperature, $\theta$, is – 12.7 K and – 11.6 K for the magnetic field parallel and perpendicular to the $c$-axis, respectively. The $\mu_{eff}$ values are close to the theoretical value of Hund's rule for the ground state of the isolated Ho$^{3+}$ ions ($\mu_{eff}$ = 10.6$\mu_B$) [27,29]. It is therefore known that the incommensurate magnetic order is dominant during phase II. However, the commensurate magnetic order is evolves with a decreasing temperature, which also accompanies the elastic softening. In phase III, the quadrupolar ordering and magnetic dipole ordering coexist and the lattice distortion is also stabilized as a consequence of the strong quadrupole-strain interactions [27].

Figure 2 shows the isothermal magnetization curves at various temperatures in the range of 2 K ≤ $T$ ≤ 50 K with a field applied along the $c$-axis for HoB$_4$. The isothermal data at $T$ = 2 K shows two meta-magnetic transitions at $H \approx$ 20 kOe and 35 kOe. The magnetic moment of the Ho$^{3+}$ ion is 6.6$\mu_B$ at $H$ =

50 kOe, which is smaller than the maximum moment of the $Ho^{3+}$ ion ($10.6\mu_B$). This indicates that the canted antiferromagnetic moments by the coupling between the quadrupolar moments and the magnetic dipole moments undergo a field-induced phase transition with an increasing field [27,31]. Typical paramagnetic behaviour is observed at $T = 50$ K.

The magnetic entropy change, $\Delta S_M$, can be estimated from the Maxwell equation in the approximated form

$$\Delta S_M (T, H) = \sum_i \frac{M_{i+1}(T_{i+1}, H_i) - M_i(T_i, H_i)}{T_{i+1} - T_i} \Delta H_i$$

where $M_{i+1}$ and $M_i$ are the experimentally measured values at temperatures $T_{i+1}$ and $T_i$, respectively, in the magnetic field, $H_i$. The temperature dependence of the magnetic entropy change of $HoB_4$ is calculated using isothermal magnetization data with a magnetic field applied along the *c*-axis and in the *ab* plane. These are plotted in Figures 3(a) and (b), respectively. For the field applied along the *c*-axis, a large positive entropy change is observed below $T = T_{N1}$ with a maximum value of 22.7 J/kg·K at $T = T_{N2}$ with $\Delta H = 25$ kOe, as shown in Figure 3(a). The maximum $\Delta S_M$ decreases with a further increase of the magnetic field. The entropy change at $T = T_{N2}$ with $\Delta H = 50$ kOe is relatively small compared to the value with $\Delta H = 25$ kOe. The entropy change near $T = T_{N1}$, which is negative, increases monotonically as the field increases, yielding $\Delta S_M = -15.9$ J/kg·K at $T = 8$ K with $\Delta H = 50$ kOe, which is a typical characteristic of the conventional MCE. For the field applied along the *ab* plane, a positive entropy change is observed with a maximum value of 10.75 J/kg·K near $T = T_{N2}$ with $\Delta H = 25$ kOe. At $T = T_{N1}$, the entropy change becomes negative with a value of $-9.8$ J/kg·K with $\Delta H = 50$ kOe.

Figure 4 shows the temperature-dependence of magnetization divided by the applied magnetic field, parallel and perpendicular to the *c*-axis, namely, $M(T)/H$ with $H = 10$ kOe, for a single crystal of $DyB_4$. Similar to the $HoB_4$ case, there are two successive magnetic transitions at $T_{N2} = 13.0$ K and $T_{N1} = 20.5$ K for an applied magnetic field parallel and perpendicular to the *c*-axis, respectively [30]. It is determined that the origins of these two transitions are quite similar to those in HoB$_4$ and are responsible for the magnetic order and quadrupolar order. However, the types of their orderings in DyB$_4$ are distinct from the ones in HoB$_4$. At $T = T_{N1}$, collinear antiferromagnetic ordering is developed along the *c*-axis, while quadrupolar ordering is developed at $T = T_{N2}$, which accompanies the structural distortion and magnetic order in both the ab-plane and the *c*-axis [28].

The isothermal magnetization curves at various temperatures in the range of 2 K ≤ $T$ ≤ 50 K with a field applied along the *c*-axis for DyB$_4$ are plotted in Figure 5. The magnetizations at $T = 2$ K and 5 K undergo field-induced transitions near $H \approx 45$ kOe and the magnetic moment of a Dy$^{3+}$ ion is observed to be 3.8 μ$_B$ at $H = 50$ kOe [32]. Because the theoretical value of the Dy$^{3+}$ ion moment is 10.6 μ$_B$, the transition is likely to be a meta-magnetic transition in the orbital ordered state, corresponding to the first meta-magnetic transition at $H = 25$ kOe in HoB$_4$. The isothermal curves are found to follow paramagnetic behaviour at $T = 20$ K and 50 K.

The temperature dependences of the magnetic entropy change for DyB$_4$ is calculated using the isothermal magnetization data for a magnetic field applied along the *c*-axis (Figure 5) and in the *ab* plane (not shown) and are plotted in Figures 6(a) and (b). For a field applied along the *c*-axis, a large positive entropy change is observed below $T = T_{N2}$ with a maximum value of 19.6 J/kg·K at $T = T_{N2}$ with $\Delta H = 50$ kOe, as shown in Figure 6(a). The entropy change near $T = T_{N1}$ is relatively small and negative. On the other hand, there is no significant magnetic entropy change for a field applied in the *ab* plane with a positive ΔS = 1.80 J/kg·K near $T = T_{N2}$ and a negative ΔS = - 3.40 J/kg·K near $T = T_{N1}$ with $\Delta H = 50$ kOe.

It is quite interesting that the maximum entropy change (Δ$S_M$) occurs at $T = T_{N2}$ with the magnetic field of the meta-magnetic transition for both HoB$_4$ and DyB$_4$. Because the meta-magnetic transition is a field-induced spin reorientation, which is strongly coupled with the quadrupole moment, a steep increase in Δ$S_M$ near $T = T_{N2}$ with a maximum Δ$S_M$ at $T = T_{N2}$ along the *c*-axis should be associated with the orbital ordering degeneracy release. In particular, the entropy change of HoB$_4$ at $T_{N2}$= 5.7 K increases as the field increases, reaching its maximum at $\Delta H$ = 25 kOe and then decreasing with a further increase in the field. Similarly, the entropy change of DyB$_4$ is maximized at $T_{N2}$ = 13.0 K and the critical field $H$ = 45 kOe, which is expected to be suppressed with a further increase of the field. This field dependence of the entropy change is quite unusual compared to the conventional behaviour, which originates purely from magnetic moments (unprecedented behaviour to our knowledge). Thus, the large magnetocaloric effect of both HoB$_4$ and DyB$_4$ near $T = T_{N2}$ is a consequence of the magnetic moment reorientation, strongly coupled with quadrupolar ordering in the presence of strong spin-orbit coupling.

To understand such behaviour of a large *inverse* MCE, it is important to note that the system is geometrically frustrated and metallic. Geometrical frustration and thus competing exchange interactions are crucial for inducing two successive magnetic transitions [33,34]. In addition, the magnetic moment and quadrupole moment are strongly coupled and mediated via itinerant electrons in this metallic system. This leads to a sudden change in the magnetization near the critical field, especially at transition temperature $T_{N2}$, where the quadrupole moment is being developed. Within Landau theory, one can qualitatively understand this phenomena taking three order parameters into account, namely, the antiferromagnetic order ($M_s$), the ferroquadrupolar order ($Q$) and the uniform magnetization ($M_u$).

$$F(M_s, M_u, Q, H) = u_s\ M_s^4 + r_s\ M_s^2 + u_Q\ Q^4 + r_Q\ Q^2 + u_u\ M_u^4 + r_u\ M_u^2 + g(H)M_u$$

$$F_{int}(M_s, M_u, Q) = w\ M_s^2 Q^2 + v\ M_u^2 Q^2 + l\ M_u^2\ M_s^2$$

$F(M_s, M_u, Q, H)$ represents the Landau free energy with mass and quartic interaction terms for $M_s$, $M_u$ and $Q$. In terms of the uniform magnetization, $M_u$ has an additional term $g(H)M_u$ due to the dominant coupling with a magnetic field $H$. Here, the function $g(H)$ takes the magnetization jump at a critical field $H_c$ into account, thus $g(H) = c\ (\ Tanh\left[\frac{H_c-H}{T}\right] - Tanh\left[\frac{H_c}{T}\right]\ )$, where c is some constant and $T$ is the temperature. $F_{int}(M_s, M_u, Q)$ represents the Landau free energy for interactions between the order parameters at quartic levels.

The consecutive phase transitions can be explained by taking the mass terms $r_s = \frac{T - T_M}{T_M}$ and $r_Q = \frac{T - T_Q}{T_Q}$, where the system stabilizes the antiferromagnetic order and ferroquadrupolar order at $T_M$ and $T_Q$, respectively ($T_Q < T_M$). Of course, these transition temperatures can be shifted in the presence of the interaction term $w\ M_s^2 Q^2$ in $F_{int}\ (M_s, M_u, Q)$. When the magnetic field is applied, the uniform magnetization $M_u$ is being developed. In this case, the interaction terms in $F_{int}\ (M_s, M_u, Q)$ lead the magnitude of $M_u$ to change non-monotonically near the transition temperatures $T_M$ and $T_Q$. We set the parameters $u_s$, $u_Q$ and $u_u$ to be all positive for stability of the order parameters, and $r_u > 0$ when the ferromagnetism is absent without a field. For the interaction terms in $F_{int}\ (M_s, M_u, Q)$, we set $w < 0$, $v > 0$ and $l > 0$. The inset of Figure 6(a) shows the calculated entropy change as a function of temperature with different magnetic field strengths. When the field strength approaches the critical field $H_c$, the positive entropy change is maximized near the transition temperature of the quadrupolar order ($T_Q$). Such a positive entropy change exists only below the magnetic ordering temperature ($T_M$), with the negative entropy change shown for $T > T_M$, as expected for a conventional MCE. This behaviour is qualitatively in good agreement with the entropy change observed in rare-earth tetraborides $DyB_4$ and $HoB_4$.

To observe the correlation of the field dependence of the entropy change with the spin-orbit interaction strength, a $Dy_{0.5}Ho_{0.5}B_4$ single crystalline specimen is synthesized. Figure 7(a) shows the temperature dependence of the magnetization with a magnetic field of $H = 10$ kOe parallel and perpendicular to the $c$-axis. Two successive transitions at $T_{N1} = 13.8$ K and $T_{N2} = 9.5$ K are observed, which are in-between those of $DyB_4$ and $HoB_4$, as expected. The entropy change for the field applied parallel to the $c$-axis is calculated from the isothermal magnetization data (see supplementary data) and is plotted in Figure 7(b) as a function of temperature with various fields of $H = 10, 20, 30, 40,$ and $50$ kOe. The maximum positive $\Delta S_M$ (= 19.0 J/kg·K) is found near $T = T_{N2}$ with $\Delta H = 40$ kOe. This provides clear evidence that the critical field for the maximum entropy change is correlated with the quadrupole coupling strength and the magnetic ground state.

The data for three compounds ($DyB_4$, $Dy_{0.5}Ho_{0.5}B_4$ and $HoB_4$) are summarized in Table 1. It is believed that the large positive entropy change near $T = T_{N2}$ is not due to a simple antiferromagnetic transition at $T = T_{N1}$, but instead due to the strong correlation between the quadrupole ordering and magnetic moment. There is no significant variation in the maximum $\Delta S_M$ values of the three compounds, indicating that the accumulated magnetic degeneracies due to a quadrupolar interaction with a magnetic dipole moment are similar for the three compounds. The critical field, i.e., the magnetic field at which the maximum entropy change is found, decreases gradually to be 50 kOe for $DyB_4$, 40 kOe for $Dy_{0.5}Ho_{0.5}B_4$ and 25 kOe for $HoB_4$. Because these fields are close to those for the field-induced meta-magnetic transitions, the critical field should be correlated with the coupling strength between the magnetic and quadrupole moments. Thus, with a further increase in the magnetic field above the critical field, the effects of quadrupolar ordering on the magnetic entropy change, $\Delta S_M$, would decrease, contrary to the conventional MCE.

## IV. CONCLUSIONS

We have systematically investigated the magnetic entropy change, i.e., the MCE and their correlation with multipolar phase transitions, for three compounds: $DyB_4$, $Dy_{0.5}Ho_{0.5}B_4$ and $HoB_4$. These three compounds exhibit common features of double phase transitions, where the magnetic order is developed at $T_{N1}$ and the quadrupolar order is developed at $T_{N2}$ ($T_{N1}$ ~20.5 K, 13.8 K, 7 K and $T_{N2}$ ~ 13.0 K, 9.5 K, 5.7 K for $DyB_4$, $Dy_{0.5}Ho_{0.5}B_4$ and $HoB_4$, respectively). Surprisingly, an enormous positive entropy change, i.e., an *inverse* MCE, is observed near $T = T_{N2}$ as the magnetic field approaches the critical field of the meta-magnetic transitions ($\approx$ 50 kOe, 40 kOe, and 25 kOe for $DyB_4$, $Dy_{0.5}Ho_{0.5}B_4$ and $HoB_4$, respectively). The maximum $\Delta S_M$ values are estimated as 19.6 J/kg·K, 19.0 J/kg·K and 22.7 J/kg·K for $DyB_4$, $Dy_{0.5}Ho_{0.5}B_4$ and $HoB_4$, respectively. While a conventional magnetic entropy change is observed at $T = T_{N1}$, as is expected for the conventional MCE, a giant *inverse* MCE observed at $T = T_{N2}$ is quite striking. Furthermore, these maximum values of $\Delta S_M$ for $Dy_{1-x}Ho_xB_4$ are very close to the largest values reported among the magnetocaloric materials in the low-field region ($H \leq 20$ kOe) and the largest reported for antiferromagnetic compounds.

Such an exotic inverse MCE is very unique, originating from an interplay of a strong spin-orbit coupling and geometric frustration. Strong spin-orbit coupling and crystal field splitting lead to the formation of a pseudo-quartet for low lying states and allows for multipolar degrees of freedom, including both magnetic dipoles (linear in total angular momentum $J$) and quadrupoles (quadratic in $J$). In rare-earth tetraborides, such multipolar degrees of freedom lie in a geometrically frustrated Shastry-Sutherland lattice and interact with one another via itinerant electrons. This induces consecutive double phase transitions in the presence of strong coupling between magnetic dipoles and quadrupoles. Thus, the magnetization change as a function of temperature is expected to be enhanced at $T = T_{N2}$ with the critical magnetic field of the meta-magnetic transition, resulting in a large *inverse* MCE. This new mechanism opens a potential pathway to understanding the physical origin of a large *inverse* MCE in rare earth tetraborides. In addition, it enables us to enhance the controllability of the MCE with several parameters and applications for other candidate materials with strong spin-orbit coupling.

Acknowledgements

This work was supported by LG Electronics Inc. and the National Research Foundation of Korea (NRF), funded by the Ministry of Science, ICT & Future Planning (No. NRF-2015M3A9B8032703 and No.NRF-2017R1A2B2008538). S.B.L is supported by the KAIST start-up and the National Research Foundation Grant (NRF-2017R1A2B4008097).

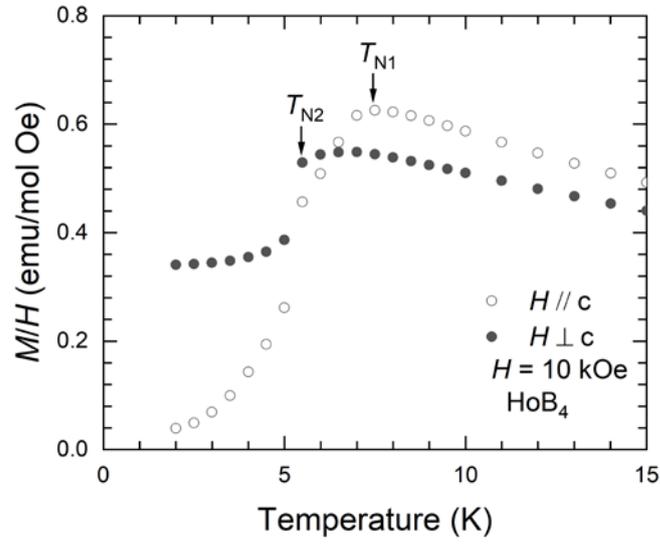

Figure 1. Temperature-dependent magnetization divided by an applied magnetic field, $H$ = 10 kOe, parallel and perpendicular to the $c$-axis for HoB$_4$.

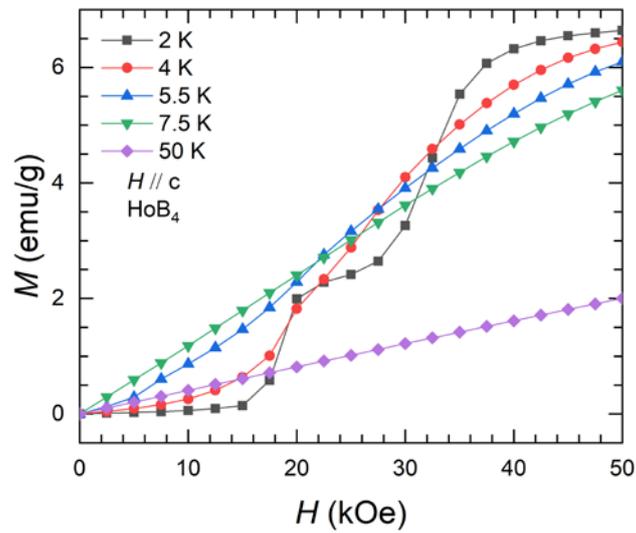

Figure 2. Magnetic field dependence of the isothermal magnetization at different temperatures in a range of 2 K ≤ $T$ ≤ 50 K along the $c$-axis for HoB$_4$.

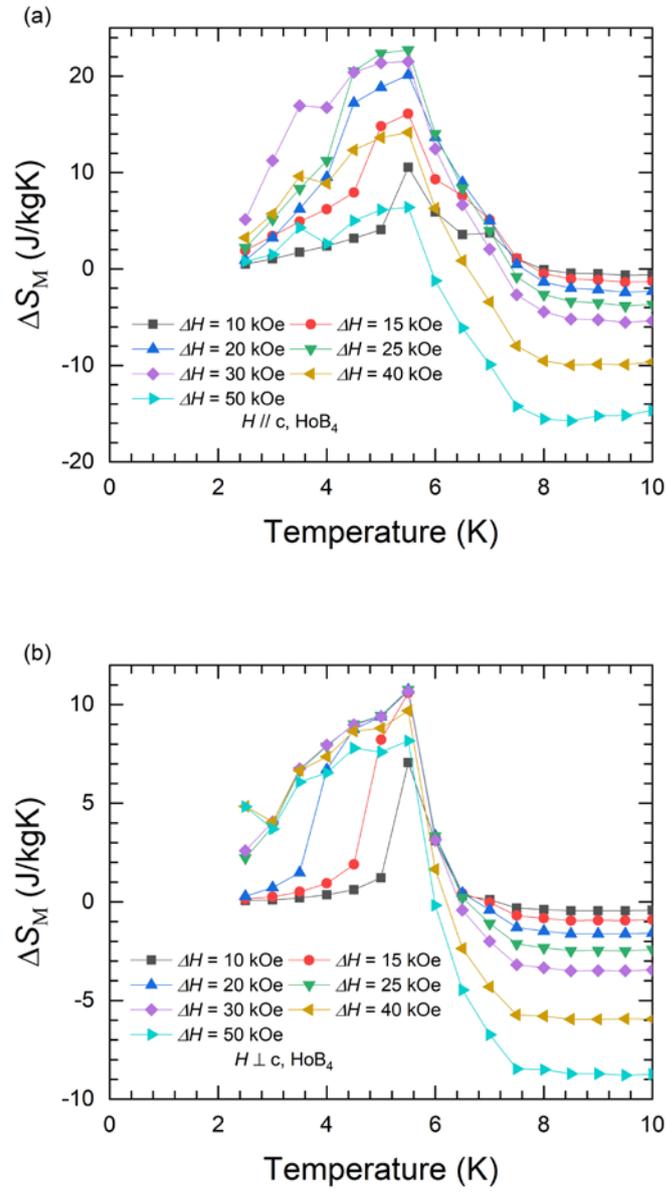

Figure 3. Temperature dependence of the magnetic entropy change in HoB$_4$ under different magnetic fields of $\Delta H$ = 10, 15, 20, 25, 30, 40 and 50 kOe, applied along the *c*-axis (a) and *ab* plane (b).

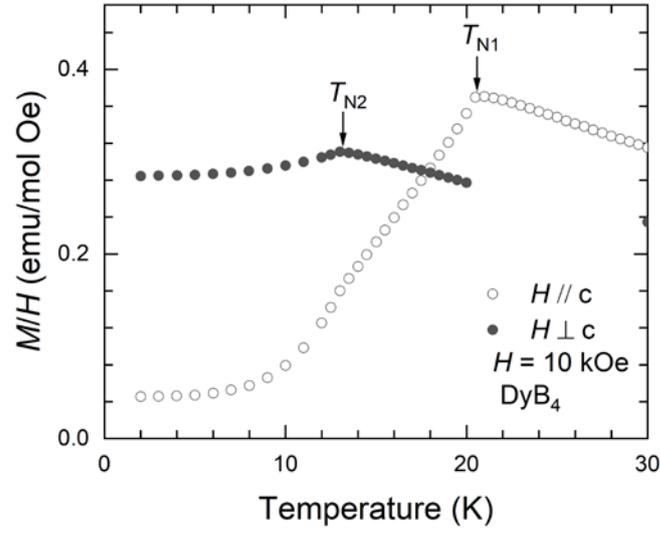

Figure 4. Temperature-dependent magnetization divided by an applied magnetic field, $H$ = 10 kOe, parallel and perpendicular to the $c$-axis for DyB$_4$.

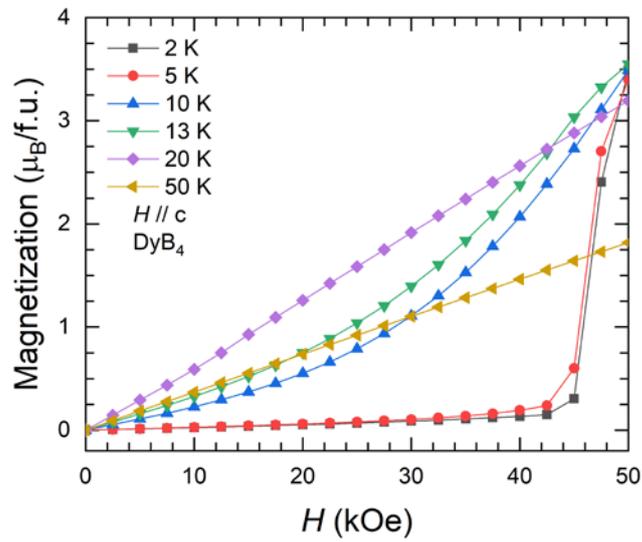

Figure 5. Magnetic field dependence of the isothermal magnetization at different temperatures in a range of 2 K ≤ $T$ ≤ 50 K along the $c$-axis for DyB$_4$.

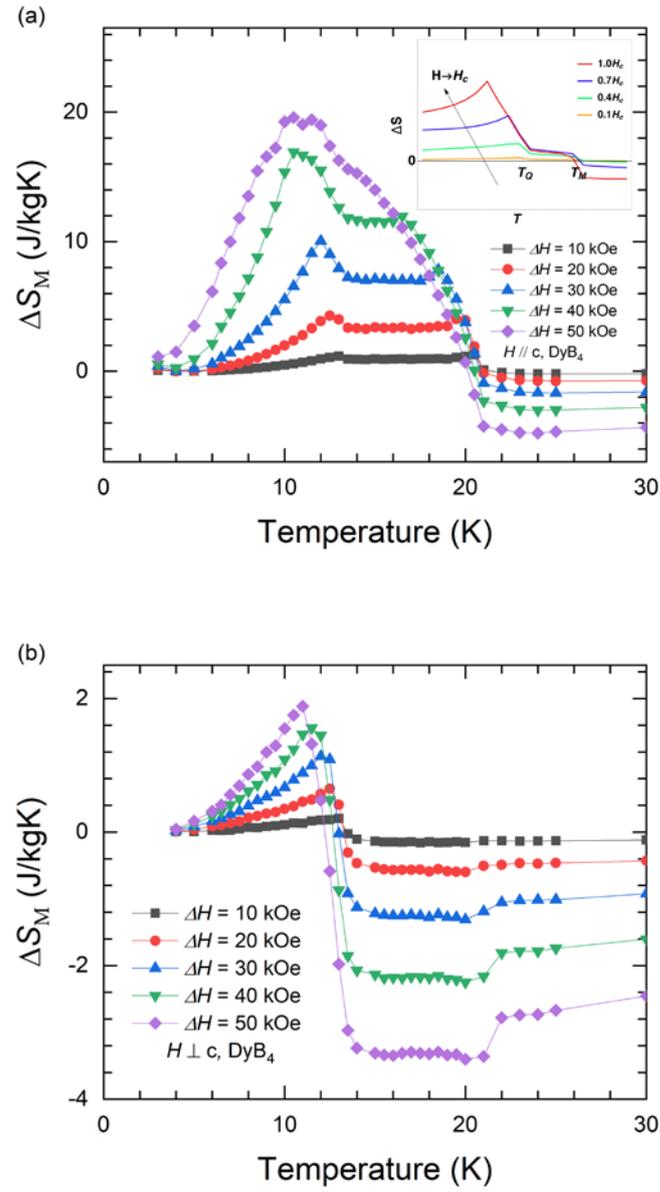

Figure 6. Temperature dependence of the magnetic entropy change of DyB$_4$ under different magnetic fields of $\Delta H$ = 10, 20, 30, 40 and 50 kOe, applied along the *c*-axis (a) and *ab* plane (b). Inset of (a): Theoretical calculation of the entropy change based on the Landau free energy. (See the main text for more details.)

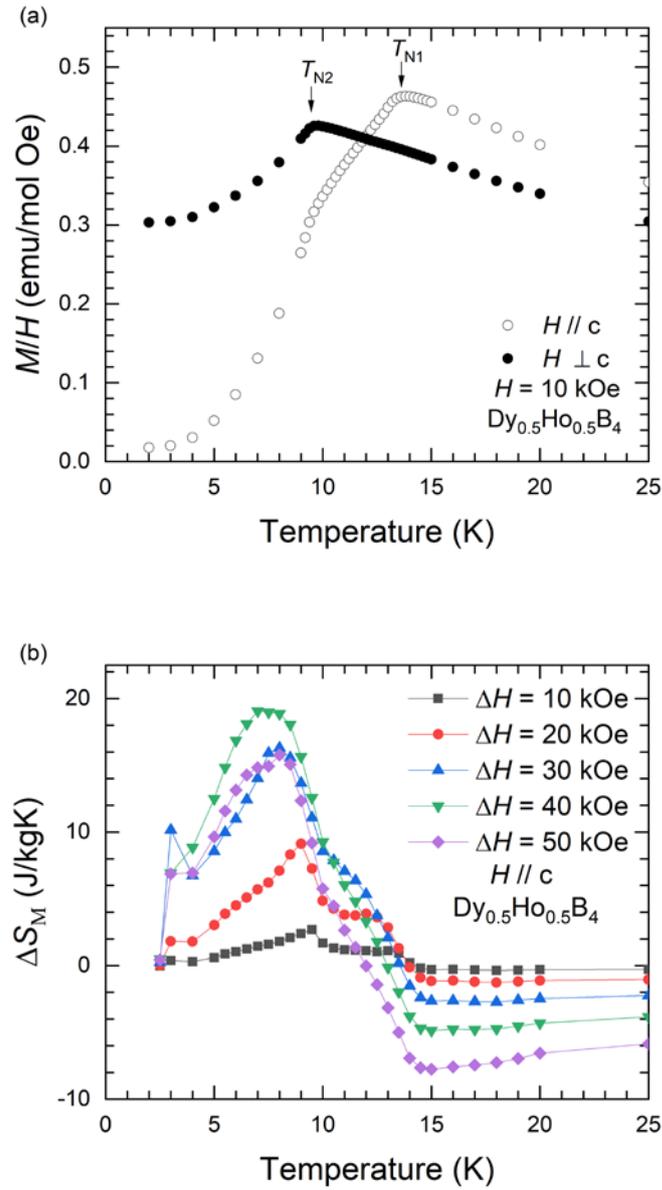

Figure 7. (a) Temperature-dependent magnetization divided by a magnetic field, $H = 10$ kOe, applied parallel and perpendicular to the $c$-axis for $Dy_{0.5}Ho_{0.5}B_4$. (b) Temperature-dependence of the magnetic entropy change in $Dy_{0.5}Ho_{0.5}B_4$ under different magnetic fields of $\Delta H = 10, 20, 30, 40$ and $50$ kOe, applied along the $c$-axis.

|  | $T_{N2}$ (K) | $T_{N1}$ (K) | $T_{N1}$ - $T_{N2}$ | $+\Delta S_M$ (J/kgK) | $\Delta H_{max}$ (T) |
|---|---|---|---|---|---|
| DyB$_4$ | 13 | 20.1 | 7.1 | 19.6 | ≈5 |
| Dy$_{0.5}$Ho$_{0.5}$B$_4$ | 9.5 | 13.8 | 4.3 | 19.0 | 4 |
| HoB$_4$ | 5.5 | 7.5 | 2 | 22.72 | 2.5 |

Table 1. Experimental data for single crystalline samples of Ho$_{1-x}$Dy$_x$B4. $T_{N2}$: quadrupolar ordering temperature, $T_{N1}$: antiferromagnetic transition temperature, $\Delta S_M$: maximum entropy change along the $c$-axis, and $\Delta H_{max}$: field for maximum magnetic entropy change.

Supplementary information

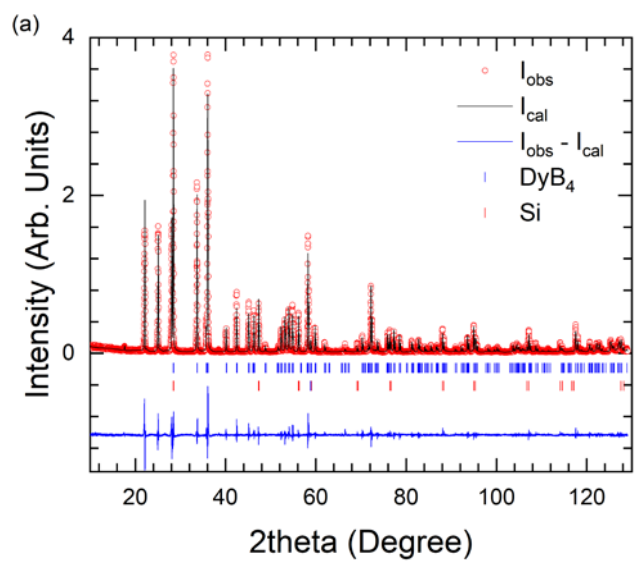

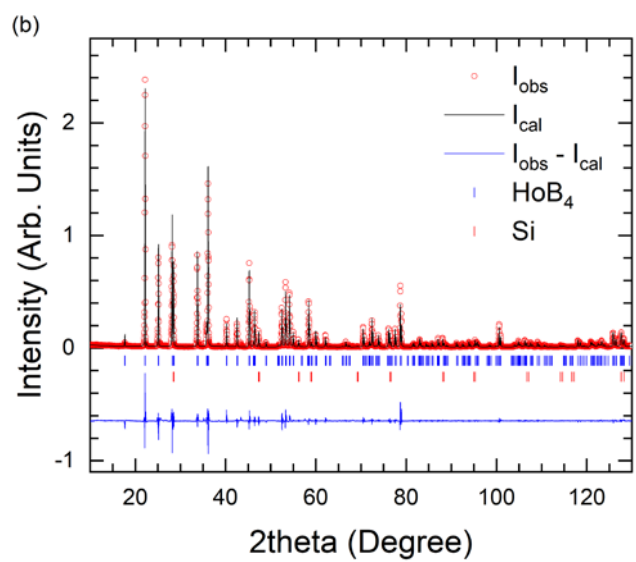

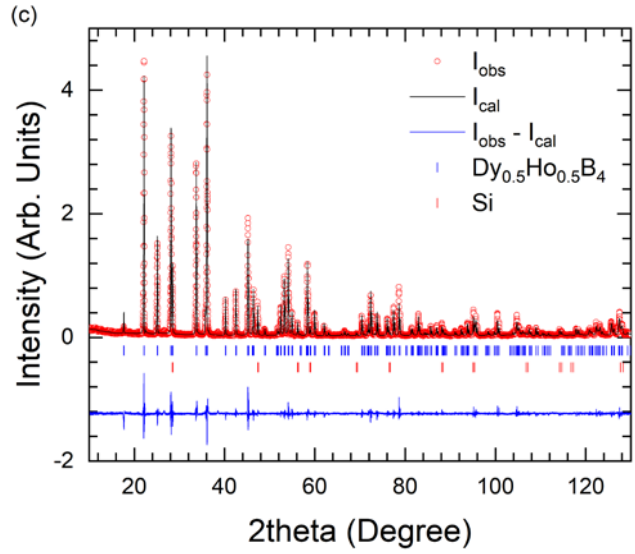

Supplementary Figure 1: Powder X-ray diffraction patterns of DyB$_4$, HoB$_4$ and Dy$_{0.5}$Ho$_{0.5}$B$_4$. The black solid line is the theoretical diffraction pattern, which is obtained using FULLPROF. The blue and red vertical bars are the Bragg peak positions of the samples and of Si, respectively. The lowest curves represent the difference between the experimental and calculated intensities.

|  | a (Å) | b (Å) | c (Å) | $T_{N2}$ (K) | $T_{N1}$ (K) |
|---|---|---|---|---|---|
| DyB$_4$ | 7.099(5) | 7.099(5) | 4.015(9) | 13 | 20.1 |
| Dy$_{0.5}$Ho$_{0.5}$B$_4$ | 7.088(4) | 7.088(4) | 4.008(7) | 9.5 | 13.8 |
| HoB$_4$ | 7.083(8) | 7.083(8) | 4.005(1) | 5.5 | 7.5 |

Supplementary Table 1: Refined lattice parameters of Ho$_{1-x}$Dy$_x$B$_4$ ($x$ = 0.0, 0.1, and 1.0).

Supplementary Note 1.

The powder X-ray diffraction patterns are collected for the pulverized single crystals. The patterns show a single phase without any noticeable impurities. The crystal structure is consistent with the tetragonal symmetry of the ThB4-type structure with the space group *P4/mbm* (#127) at room temperature. The lattice constants vary, and the volume decreases with an increasing Ho concentration in Ho$_{1-x}$Dy$_x$B$_4$. The lattice parameters and volume follow Vegard's law. The antiferromagnetic transition temperature is also observed to systematically decrease as the Ho concentration increases, which indicates a uniform distribution of both Ho and Dy in Ho$_{1-x}$Dy$_x$B$_4$.

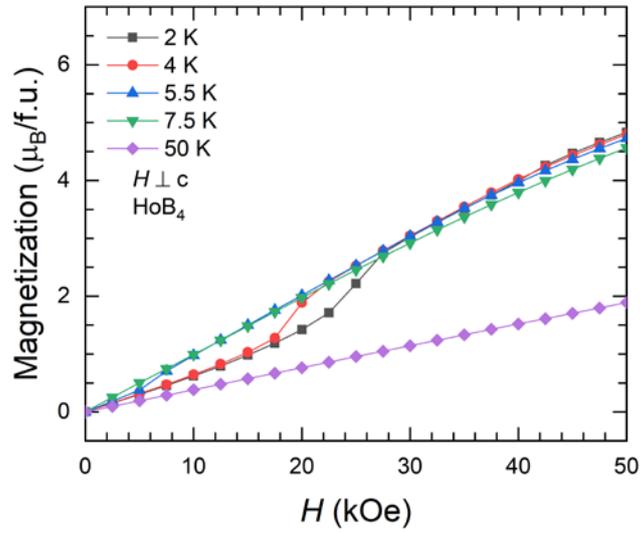

Supplementary Figure 2. Magnetic field dependence of the isothermal magnetization at different temperatures in a range of 2 K ≤ $T$ ≤ 50 K along the *ab*-plane for HoB$_4$.

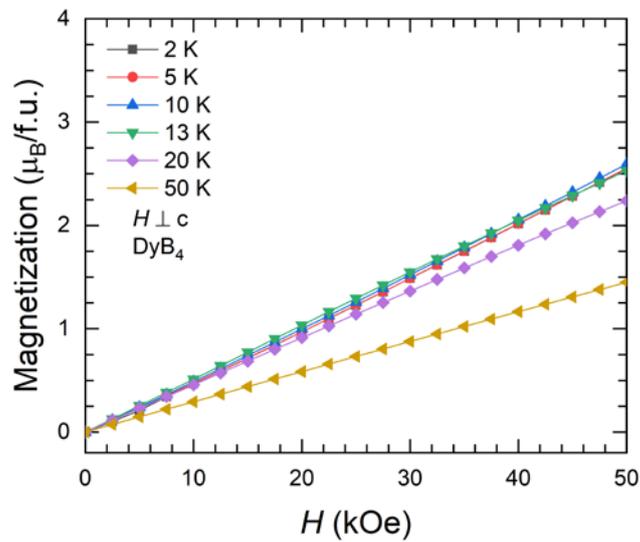

Supplementary Figure 3. Magnetic field dependence of the isothermal magnetization at different temperatures in a range of 2 K ≤ $T$ ≤ 50 K along the *ab*-plane for DyB$_4$.

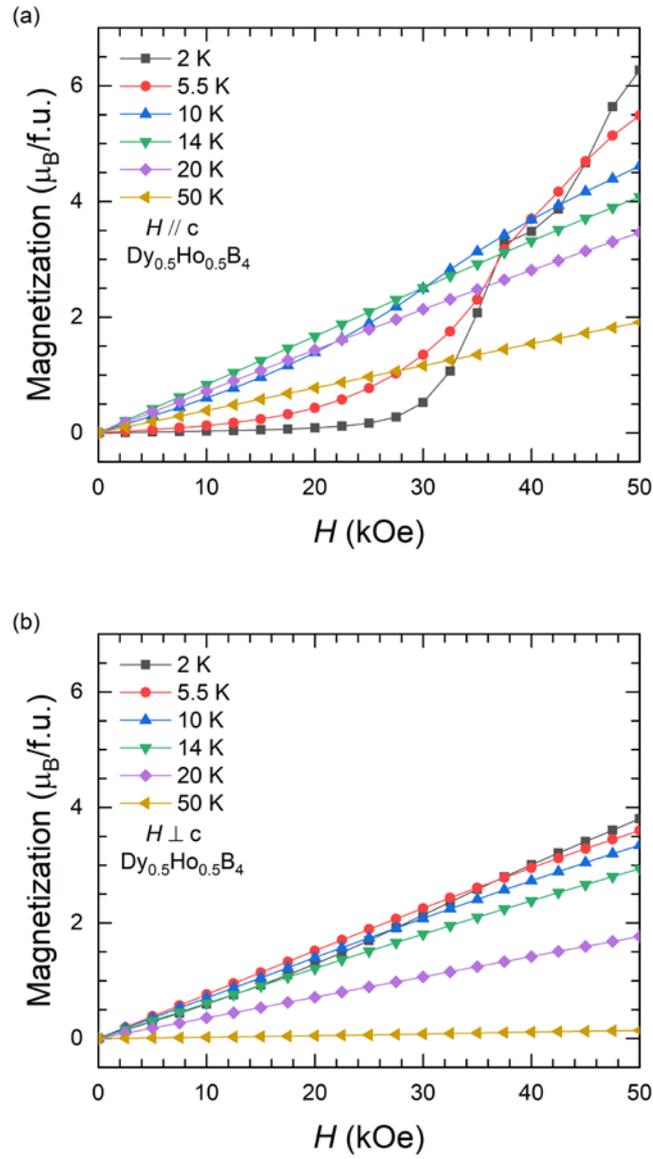

Supplementary Figure 4: Magnetic field dependence of the isothermal magnetization at different temperatures in a range of 2 K ≤ $T$ ≤ 50 K along the $c$-axis (a) and $ab$ plane (b) for $Dy_{0.5}Ho_{0.5}B_4$.

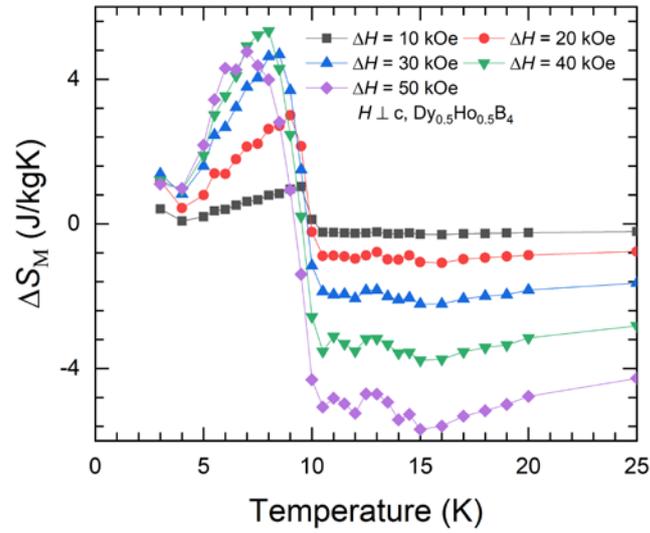

Supplementary Figure 5: Temperature-dependence of the magnetic entropy change of $Dy_{0.5}Ho_{0.5}B_4$ with different magnetic fields of $\Delta H$ = 10, 20, 30, 40 and 50 kOe along the *ab* plane.

Supplementary Reference